\documentstyle[12pt,epsf]{article}

\begin{document}

\begin{titlepage}

\begin{flushright}
CERN-TH/97-342\\
hep-ph/9712224
\end{flushright}

\vspace{2.0cm}
\begin{center}
\boldmath
\Large\bf
Rescattering Effects, Isospin Relations and\\
Electroweak Penguins in $B\to\pi K$ Decays
\unboldmath
\end{center}

\vspace{0.5cm}
\begin{center}
Matthias Neubert\\[0.1cm]
{\sl Theory Division, CERN, CH-1211 Geneva 23, Switzerland}
\end{center}

\vspace{1.5cm}
\begin{abstract}
\vspace{0.2cm}\noindent
It is argued that, in the presence of soft final-state interactions,
the diagrammatic amplitude approach adopted in many analyses of
hadronic $B$ decays into light mesons can be misleading when used to
deduce the unimportance of certain decay topologies. With the example
of $B\to\pi K$ decays, it is shown that the neglect of so-called
annihilation and colour-suppressed amplitudes (including electroweak
penguins), as well as penguin contributions involving an up-quark loop,
is not justified. The implications for the Fleischer--Mannel bound on
the angle $\gamma$ of the unitarity triangle, and for the CP asymmetry
in the decays $B^\pm\to\pi^\pm K^0$, are pointed out.
\end{abstract}

\vspace{1.0cm}
\centerline{(Submitted to Physics Letters B)}

\vspace{4.0cm}
\noindent
CERN-TH/97-342\\
December 1997

\end{titlepage}

The study of CP violation in the rare decays of $B$ mesons is the main
target of present and future ``$B$ factories''. It is hoped that this
will shed light on the origin of CP violation, which may lie outside
the standard model of strong and electroweak interactions. Whereas at
present only a single measurement of a CP-violating asymmetry exists
(the quantity $\epsilon_K$ in $K$ decays), the measurements of several
CP asymmetries in $B$ decays will make it possible to test whether the
CKM mechanism of CP violation is sufficient to account for the data, or
whether additional sources of CP violation are required (for some
excellent recent reviews, see Refs.~\cite{Flei,rev}). In the latter
case, this would directly point towards physics beyond the standard
model.

In order to achieve this goal, it is necessary that the theoretical
calculations of CP-violating observables in terms of standard model
parameters are, at least to a large extent, free of hadronic
uncertainties. This can be achieved, for instance, by measuring
time-dependent asymmetries in the decays of neutral $B$ mesons into CP
eigenstates, such as $B\to J/\psi\,K_S$. In many other
cases, however, there are nontrivial strong-interactions effects
affecting the CP asymmetries. In the absence of a reliable theoretical
approach to calculate these effects, strategies have been developed
that exploit the isospin symmetry of the strong interactions, or its
approximate SU(3) flavour symmetry, to derive relations between
various decay amplitudes, which can be used to eliminate hadronic
uncertainties (see Refs.~\cite{zepp}--\cite{Gron} for some early
applications of this approach). A comprehensive review of these methods
can be found in Ref.~\cite{Flei}.

In this note, we question the theoretical approximations underlying
some of these analyses, which need to rely on ``plausible'' dynamical
assumptions such as the neglect of colour-suppressed or annihilation
topologies. To be specific, we consider the decay amplitudes for the
various $B\to\pi K$ modes and analyse the relations among them imposed
by isospin symmetry. The effective weak Hamiltonian governing these
transitions has the structure \cite{Heff}
\begin{equation}
   {\cal H}_{\rm eff}
   = \sum_{i=1,2}\,\Big[ \lambda_u\,Q_i^u + \lambda_c\,Q_i^c \Big]
   + \lambda_t\,\sum_{i=3}^{10}\,Q_i + \mbox{h.c.} \,,
\end{equation}
where $\lambda_q=V_{qs} V_{qb}^*$ are products of elements of the CKM
matrix, satisfying the unitarity relation
$\lambda_u+\lambda_c+\lambda_t=0$, and $Q_i$ represent the products of
local four-quark operators with short-distance coefficient functions.
Relevant for our purposes are only the isospin quantum numbers of these
operators: the current--current operators $Q_{1,2}^u\sim\bar b s\bar u
u$ have components with $\Delta I=0$ and $\Delta I=1$; the
current--current operators $Q_{1,2}^c\sim\bar b s\bar c c$, as well as
the QCD penguin operators $Q_{3,\dots,6}\sim\bar b s\sum\bar q q$, have
$\Delta I=0$; the electroweak penguin operators $Q_{7,\dots,10}\sim\bar
b s\sum e_q\bar q q$, where $e_q$ are the electric charges of the
quarks, have components with $\Delta I=0$ and $\Delta I=1$. Thus, we
may write ${\cal H}_{\rm eff} = {\cal H}_{\Delta I=0} + {\cal
H}_{\Delta I=1}$ with
\begin{eqnarray}
   {\cal H}_{\Delta I=0} &=& \sum_{i=1,2}\,\left[
    \frac{\lambda_u}{2}\,(Q_i^u + Q_i^d) + \lambda_c\,Q_i^c \right]
    + \lambda_t\,\sum_{i=3}^{10}\,Q_i - \lambda_t\,\sum_{i=7}^{10}\,
    Q_i^{\Delta I=1} + \mbox{h.c.} \,, \nonumber\\
   {\cal H}_{\Delta I=1} &=& \sum_{i=1,2}\,\frac{\lambda_u}{2}\,
    (Q_i^u - Q_i^d) + \lambda_t\,\sum_{i=7}^{10}\,Q_i^{\Delta I=1}
    + \mbox{h.c.} \,,
\label{Hdecomp}
\end{eqnarray}
where the $\Delta I=1$ components of the electroweak penguin operators
are defined as $Q_{7,\dots,10}^{\Delta I=1}\sim\frac 12\bar b s(\bar u
u-\bar d d)$. Taking into account that the initial $B$-meson state has
$I=\frac 12$, whereas the final states $(\pi K)$ can be decomposed into
states with $I=\frac 12$ and $I=\frac 32$, the Wigner--Eckart theorem
implies that the physical decay amplitudes can be described in terms of
three isospin amplitudes, which are defined as \cite{MGr,NQ}
\begin{eqnarray}
   A_{3/2} &=& \textstyle \sqrt{\frac 13}\,
    \langle\frac 32,\pm\frac 12| {\cal H}_{\Delta I=1}
    |\frac 12,\pm\frac 12\rangle \,, \nonumber\\
   A_{1/2} &=& \textstyle \pm\sqrt{\frac 23}\,
    \langle\frac 12,\pm\frac 12| {\cal H}_{\Delta I=1}
    |\frac 12,\pm\frac 12\rangle \,, \nonumber\\
   B_{1/2} &=& \textstyle \sqrt{\frac 23}\,
    \langle\frac 12,\pm\frac 12| {\cal H}_{\Delta I=0}
    |\frac 12,\pm\frac 12\rangle \,.
\label{isoampl}
\end{eqnarray}
{}From the decomposition of the effective Hamiltonian in
(\ref{Hdecomp}) it is obvious which operator matrix elements and weak
phases enter the various isospin amplitudes. The resulting expressions
for the physical $B\to\pi K$ decay amplitudes are given by
\begin{eqnarray}
   {\cal A}_{-+} &=& {\cal A}(B^0\to\pi^- K^+)
    = A_{3/2} + A_{1/2} - B_{1/2} \,, \nonumber\\
   {\cal A}_{+0} &=& {\cal A}(B^+\to\pi^+ K^0)
    = A_{3/2} + A_{1/2} + B_{1/2} \,, \nonumber\\
   {\cal A}_{00} &=& \sqrt{2}\,{\cal A}(B^0\to\pi^0 K^0)
    = 2 A_{3/2} - A_{1/2} + B_{1/2} \,, \nonumber\\
   {\cal A}_{0+} &=& \sqrt{2}\,{\cal A}(B^+\to\pi^0 K^+)
    = 2 A_{3/2} - A_{1/2} - B_{1/2} \,.
\label{ampl}
\end{eqnarray}

\begin{figure}
\epsfxsize=15cm
\centerline{\epsffile{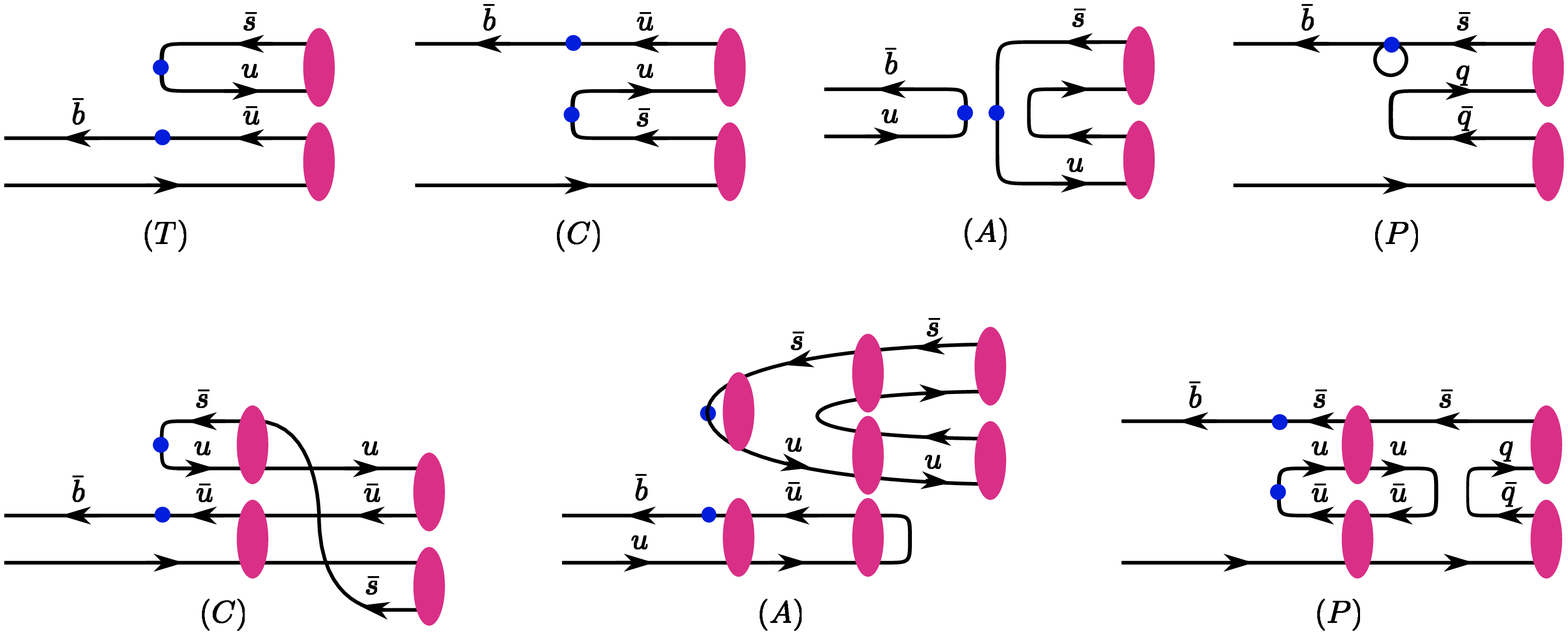}}
\centerline{\parbox{14.5cm}{\caption{\label{fig:topol}
\small\sl Flavour-flow topologies relevant to $B\to\pi K$ decays. In
the lower plots, the topologies $C$, $A$ and $P$ are redrawn as soft
final-state rescatterings from the tree amplitude. The dots indicate
the quark fields contained in the operators of the effective weak
Hamiltonian. The shaded blobs represent (intermediate) hadronic states.
}}}
\end{figure}

Instead of expressing the isospin amplitudes in terms of operator
matrix elements, many practitioners prefer to analyze the $B$ decay
amplitudes in terms of a diagrammatic notation, in which complex
amplitudes are associated with certain flavour-flow topologies
\cite{Chau,Gron}. If we neglect electroweak penguin diagrams for the
moment (we will come back to them later), the topologies relevant to
our discussion are the so-called ``tree topology'' $T$, the
``colour-suppressed tree topology'' $C$, the ``annihilation topology''
$A$, and the ``penguin topology'' $P$ shown in the upper plots in
Figure~\ref{fig:topol}. In terms of these quantities, the decay
amplitudes take the form \cite{Gron}
\begin{eqnarray}
   {\cal A}_{-+} &=& -(T+P) \,, \qquad
    {\cal A}_{00} = -C+P \,, \nonumber\\
   {\cal A}_{+0} &=& A+P \,, \hspace{1.38cm}
   {\cal A}_{0+} = -(T+C+A+P) \,,
\label{TPA}
\end{eqnarray}
while the isospin amplitudes are given by $A_{3/2}=-\frac 13(T+C)$,
$A_{1/2}=-\frac 16 T+\frac 13 C+\frac 12 A$ and $B_{1/2}=P+\frac
12(T+A)$. If electroweak penguin contributions are neglected, the
amplitudes $T$, $C$ and $A$ are proportional to $\lambda_u$, whereas
the penguin amplitude $P$ has contributions proportional to all three
$\lambda_q$, and we define $P=\lambda_u P_u + \lambda_c P_c + \lambda_t
P_t$.

The diagrammatic approach provides a redundant parametrization of the
decay amplitudes in that there are more flavour-flow topologies than
isospin amplitudes. We stress that, whereas the isospin amplitudes can
be defined in a transparent way in terms of operator matrix elements
using the decomposition (\ref{Hdecomp}), this is {\em not\/} the case
for the individual amplitudes in the diagrammatic approach. For
instance, the matrix elements of the current--current operators
$Q_{1,2}^u$ contribute to $T$, $C$, $A$ and $P_u$. Still, the approach
is perfectly legitimate in a mathematical sense, as long as one works
with {\em exact\/} expressions for the physical decay amplitudes.
However, the main virtue of the diagrammatic approach is claimed to be
the fact that it would allow one, by making ``plausible'' dynamical
assumptions, to simplify the relations between decay amplitudes. In
particular, it is usually argued that annihilation diagrams are
suppressed relative to tree diagrams by a factor of
$f_B/m_B\sim\mbox{few}\,\%$ stemming from the fact that in order to
have the quarks inside the initial $B$ meson annihilate each other
through a weak current they have to be at the same point, implying a
suppression proportional to the wave-function at the origin. This
argument is used to conclude that $|A|\ll|T|$, and hence $A$ is often
neglected. Similarly, it is argued that colour-suppressed tree diagrams
are suppressed relative to colour-allowed ones by a factor $a_2/a_1\sim
0.2$, and hence $|C|\ll|T|$. Finally, it is often assumed that penguin
diagrams are dominated by the contributions from heavy-quark loops,
whereas the contribution from the up-quark loop is neglected (i.e.\
$|\lambda_u P_u|\ll|T|$). Even the charm-penguin contribution has been
neglected in some applications; however, its importance has been
stressed recently by several authors \cite{BuFl,charmp}. With these
assumptions, the two amplitudes in (\ref{TPA}) which correspond to
processes that have been observed experimentally simplify to
\begin{equation}
   {\cal A}_{-+} \approx -(T+P) \,, \qquad
   {\cal A}_{+0} \approx P \,,
\end{equation}
where $P\approx\lambda_c (P_c-P_t)$. Since there is no nontrivial weak
phase in $\lambda_c$, one does not expect a CP asymmetry in the decays
$B^\pm\to\pi^\pm K^0$, and these equations can be used to derive a
bound on the weak phase $\gamma$ of the tree amplitude $T$. Defining
$T/P=r e^{i\gamma} e^{i\delta}$, where $\delta$ is an unknown strong
phase, one finds for the ratio of the branching ratios for the two
processes $B_d\to\pi^\mp K^\pm$ and $B^\pm\to\pi^\pm K^0$, averaged
over
CP-conjugate modes, \begin{equation}
   R = \frac{\mbox{Br}(B_d\to\pi^\mp K^\pm)}
            {\mbox{Br}(B^\pm\to\pi^\pm K^0)}
   = 1 + 2r\cos\gamma\cos\delta + r^2 \ge \sin^2\gamma \,.
\label{FMbound}
\end{equation}
This is the Fleischer--Mannel bound, which excludes a region of
parameter space around $\gamma=90^\circ$ provided that $R<1$ \cite{FM}.
Given that the current experimental value $R_{\rm exp}=0.65\pm 0.40$
\cite{CLEO} indicates that this may indeed be the case, this bound has
received a lot of attention. Its implications for CP phenomenology in
the standard model and beyond have been analyzed in
Refs.~\cite{GNPS,GR97}.

The purpose of this note is to stress that soft rescattering effects,
which have been shown to be potentially significant even in the decays
of heavy hadrons \cite{Dono}, may invalidate the assumptions about the
relative size of the amplitudes in the diagrammatic approach discussed
above, and thus may invalidate the bound in (\ref{FMbound}). Our main
point is that the topologies $C$, $A$ and $P_u$ contain contributions
corresponding to final-state rescatterings of the leading tree
amplitude $T$, as shown in the lower plots in Figure~\ref{fig:topol}.
If the (unknown) final-state phases happen to be large, it is thus {\em
natural\/} to assume that $T$, $C$, $A$ and the up-penguin $P_u$ are
all of a similar magnitude. Note, in particular, that the naive
arguments in favour of a suppression of $A$ and $C$ relative to $T$ no
longer apply. For instance, the ``wave-function suppression'' of the
annihilation amplitude is absent in the ``soft'' annihilation process
shown in the figure. Therefore, although the diagrammatic approach was
originally designed to provide a model-independent parametrization of
decay amplitudes including all strong-interaction effects, in its
practical form, in which certain dynamical approximations are adopted,
it does {\em not\/} provide an appropriate representation of the
amplitudes unless final-state rescattering effects are negligible. On
the other hand, the similar importance of the various contributions
($T$, $C$, $A$ and $P_u$) involving the CKM parameter $\lambda_u$
emerges naturally in an approach where the different isospin amplitudes
are related to operator matrix elements.

It is instructive to illustrate our point in the context of a simple
model for the weak decay amplitudes. For this purpose, we shall adopt
the generalized factorization prescription \cite{Stech} to calculate
the short-distance contributions to the matrix elements of the
current--current operators $Q_{1,2}^u$, neglect ``hard''
(short-distance) annihilation contributions and electroweak penguins,
and neglect the imaginary parts of the charm- and up-quark penguin
diagrams, which reflect long-distance contributions from physical
intermediate states. Once the short-distance contributions are
calculated, long-distance effects are accounted for by introducing
elastic rescattering phases for the two isospin channels of the
final-state mesons. In this model, the short-distance $B\to\pi K$
amplitudes are given by ${\cal A}_{-+}^{\rm SD} = -(M_T+M_P)$, ${\cal
A}_{+0}^{\rm SD} = M_P$, ${\cal A}_{00}^{\rm SD} = -M_C+M_P$, and
${\cal A}_{0+}^{\rm SD} = -(M_T+M_C+M_P)$, where $M_P$ represents the
short-distance penguin contributions, and
\begin{eqnarray}
   M_T &=& \frac{G_F}{\sqrt 2}\,\lambda_u a_1\,f_K\,(m_B^2-m_\pi^2)\,
    F_0^{B\to\pi}(m_K^2) \approx (2.5\pm 0.4)\,V_{ub}^*
    \times 10^{-6}\,{\rm GeV} \,, \nonumber\\
   M_C &=& \frac{G_F}{\sqrt 2}\,\lambda_u a_2\,f_\pi\,(m_B^2-m_K^2)\,
    F_0^{B\to K}(m_\pi^2) \approx (0.20\pm 0.06) M_T \,,
\label{MTMC}
\end{eqnarray}
are the factorized matrix elements of the current--current operators
$Q_{1,2}^u$ in the notation of Ref.~\cite{Stech}, from which we also
take the values of the hadronic form factors with conservative errors.
The ratio of the hadronic parameters $a_1$ and $a_2$ is taken as
$a_2/a_1=0.22\pm 0.05$. From a naive comparison with (\ref{TPA}), one
would conclude that $|T|=|M_T|$, $|C|=|M_C|$, $A=0$, and $|P|=|M_P|$.
These results are indeed often used to estimate the magnitudes of $T$
and $C$. This identification is {\em not\/} justified, however.
Instead, we must calculate the short-distance contributions to the
different isospin amplitudes and then account for the final-state
phases. This gives
\begin{eqnarray}
   A_{3/2} &=& -\frac 13 (M_T + M_C)\,e^{i\phi_{3/2}} \,, \nonumber\\
   A_{1/2} &=& -\frac 16 (M_T - 2 M_C)\,e^{i\phi_{1/2}} \,, \nonumber\\
   B_{1/2} &=& \left( M_P + \frac 12 M_T \right) e^{i\phi_{1/2}} \,.
\label{isofact}
\end{eqnarray}
Inserting these results into the general expressions (\ref{ampl})
yields
\begin{eqnarray}
   {\cal A}_{-+} &=& -(M_T + M_P)\,e^{i\phi_{1/2}} - X \,, \nonumber\\
   {\cal A}_{+0} &=& M_P\,e^{i\phi_{1/2}} - X \,, \nonumber\\
   {\cal A}_{00} &=& (-M_C + M_P)\,e^{i\phi_{1/2}} - 2X \,, \nonumber\\
   {\cal A}_{0+} &=& -(M_T + M_C + M_P)\,e^{i\phi_{1/2}} - 2X \,,
\label{model}
\end{eqnarray}
where
\begin{equation}
   X = \frac 13 (M_T + M_C) \Big( e^{i\phi_{3/2}} - e^{i\phi_{1/2}}
   \Big) \,.
\end{equation}
We stress that, even in a factorization approach, it is important to
include final-state rescattering effects in the way outlined above,
unless it is experimentally known that such effects are negligible
(i.e.\ that $|\phi_{3/2}-\phi_{1/2}|\ll 1$). Comparing the result
(\ref{model}) with the relations (\ref{TPA}) of the diagrammatic
approach, we now obtain
\begin{eqnarray}
   T &=& M_T\,e^{i\phi_{1/2}} + X - \Delta P \,, \hspace{1.07cm}
    A = - X - \Delta P \,, \nonumber\\
   C &=& M_C\,e^{i\phi_{1/2}} + 2 X + \Delta P \,, \qquad
    P = M_P\,e^{i\phi_{1/2}} + \Delta P \,,
\label{Xcontr}
\end{eqnarray}
where $\Delta P$ is arbitrary and cancels in the predictions for the
physical decay amplitudes. This reflects the redundancy in the
parametrization of three isospin amplitudes in terms of four
diagrammatic amplitudes. By choosing $\Delta P$ appropriately, it is
possible to redistribute the rescattering contributions between the
various amplitudes, leaving the physical decay amplitudes unchanged.
We stress that the strong phases entering the diagrammatic amplitudes
are {\em not\/} governed by the isospin of the final states fed by
these amplitudes, the reason being that the diagrammatic amplitudes are
not isospin amplitudes. For instance, we see that at least one of the
amplitudes $A$ or $P$ must contain the phase $\phi_{3/2}$, although
they both lead to final states with $I=\frac 12$ only. If we choose to
set $\Delta P=0$, for instance, we get
\begin{eqnarray}
   \frac{T}{M_T\,e^{i\phi_{1/2}}}
   &=& 1 + \frac 13 \left( 1 + \frac{M_C}{M_T} \right)
    \left( e^{i\Delta\phi} - 1 \right) \,, \nonumber\\
   \frac{C}{M_T\,e^{i\phi_{1/2}}}
   &=& \frac{M_C}{M_T} + \frac 23 \left( 1 + \frac{M_C}{M_T} \right)
    \left( e^{i\Delta\phi} - 1 \right) \,, \nonumber\\
   \frac{A}{M_T\,e^{i\phi_{1/2}}}
   &=& -\frac 13 \left( 1 + \frac{M_C}{M_T} \right)
    \left( e^{i\Delta\phi} - 1 \right) \,,
\end{eqnarray}
where $\Delta\phi=\phi_{3/2}-\phi_{1/2}$. Unless $|\Delta\phi|\ll 1$,
it is not justified to assume that $|C|\ll|T|$ or $|A|\ll|T|$. In the
presence of soft final-state interactions, there is no colour
suppression of $C$ with respect to $T$, and there is no intrinsic
smallness of the annihilation topology $A$. For a phase difference of
$45^\circ$, for instance, we find $|T|:|C|:|A|\approx 1:0.61:0.33$. If
we choose instead $\Delta P=-X$ so as to keep the annihilation
amplitude small, we would find $|T|:|C|:|\Delta P|\approx 1:0.31:0.32$.
With this choice, the up-quark penguin receives a large rescattering
contribution, which is of a similar magnitude as the tree amplitude.

Although our model is too simple to provide for a trustable calculation
of the decay amplitudes, we believe it illustrates nicely the potential
pitfalls of the diagrammatic method. A more realistic analysis would
have to include inelastic rescattering contributions \cite{Dono}. Also,
there are important contributions to the strong phase of the charm
penguin $P_c$ from rescattering through intermediate states such as
$D_s\bar D$ or $J/\psi\,K$. As a consequence, the two $I=\frac 12$
amplitudes $A_{1/2}$ and $B_{1/2}$ will, in general, acquire different
phases. To get a more realistic model, we thus modify the last relation
in (\ref{isofact}) to read
\begin{equation}
   B_{1/2} = |M_P| e^{i\phi_P} + \frac 12\,M_T\,e^{i\phi_{1/2}} \,.
\end{equation}
$M_P=\lambda_c(P_c-P_t)$ has no nontrivial weak phase, whereas $M_T$ is
proportional $e^{i\gamma}$.

Before we outline some of the implications of our results, we come back
to the discussion of electroweak penguin operators, which according to
(\ref{Hdecomp}) and (\ref{isoampl}) contribute to all three isospin
amplitudes. As far as isospin (but not SU(3) flavour) symmetry is
concerned, the $\Delta I=0$ contributions of electroweak penguins can
be absorbed into a redefinition of the top-quark penguin amplitude
$P_t$ contained in $B_{1/2}$. However, electroweak penguins with
$\Delta I=1$ cause problems, since they induce terms proportional to
$\lambda_t$ in the amplitudes $A_{3/2}$ and $A_{1/2}$. Using Fierz
identities to rewrite the current--current operators, and adopting the
notation of Ref.~\cite{Heff}, we find
\begin{eqnarray}
   {\cal H}_{\Delta I=1} &=& \frac{G_F}{2\sqrt 2}\,\Bigg\{
    \bigg[ \lambda_u C_1(\mu) - \frac 32\lambda_t C_9(\mu) \bigg]
    (\bar b_\alpha s_\alpha)_{V-A} (\bar u_\beta u_\beta
    - \bar d_\beta d_\beta)_{V-A} \nonumber\\
   &&\mbox{}+ \bigg[ \lambda_u C_2(\mu) - \frac 32\lambda_t C_{10}(\mu)
    \bigg] (\bar b_\alpha s_\beta)_{V-A} (\bar u_\beta u_\alpha
    - \bar d_\beta d_\alpha)_{V-A} + \dots \Bigg\} + \mbox{h.c.} \,,
\label{EWP}
\end{eqnarray}
where $C_i(\mu)$ are Wilson coefficients, and the ellipses represent
the contributions from the operators $Q_7$ and $Q_8$, which have a
different Dirac structure. To give an idea about the relative
importance of the various contributions, we quote the values of the
coefficients at $\mu=m_b$ (in the NDR scheme): $C_1\approx -0.185$,
$C_2\approx 1.082$, $C_7\approx -10^{-5}$, $C_8\approx 0.4\cdot
10^{-3}$, $C_9\approx -9.4\cdot 10^{-3}$, $C_{10}\approx 1.9\cdot
10^{-3}$. The values of $C_7$ and $C_8$ are so tiny that it should be a
good approximation to neglect the contributions the operators $Q_7$ and
$Q_8$. However, the other two electroweak penguins are important. Using
$\lambda_u/\lambda_t=-\lambda^2 R_b e^{i\gamma}$ with $\lambda=0.22$
and $R_b\approx 0.36$ \cite{Flei}, we find $[\lambda_u C_1-\frac
32\lambda_t C_9]\approx - |\lambda_u| (0.19 e^{i\gamma} + 0.81)$ and
$[\lambda_u C_2-\frac 32\lambda_t C_{10}]\approx |\lambda_u| (1.08
e^{i\gamma} + 0.16)$. This proves, without any assumption about
hadronic matrix elements, that electroweak penguins give an important
contribution to the $\Delta I=1$ amplitudes $A_{3/2}$ and $A_{1/2}$.
In the generalized factorization scheme, their effects can be included
by replacing the hadronic parameters $a_1$ and $a_2$ in (\ref{MTMC})
with the new values
\begin{eqnarray}
   a_1^{\rm eff} &\approx& a_1 + (0.025 a_1 - 0.740 a_2) e^{-i\gamma}
    \approx a_1 (1 - 0.14 e^{-i\gamma}) \,, \nonumber\\
   a_2^{\rm eff} &\approx& a_2 + (0.025 a_2 - 0.740 a_1) e^{-i\gamma}
    \approx a_2 (1 - 3.28 e^{-i\gamma}) \,,
\label{aieff}
\end{eqnarray}
where in the last step $a_2/a_1\approx 0.22$ has been used. To derive
(\ref{aieff}), we have only used the ansatz $a_1=C_2+\xi C_1$ and
$a_2=C_1+\xi C_2$ \cite{Stech} as well as the values of the Wilson
coefficients quoted above; the hadronic parameter $\xi$ does not enter
in this result. We observe that electroweak penguins give a large
contribution to the matrix element $M_C$ in (\ref{MTMC}), whereas their
effect on $M_T$ is moderate. Note that the notion of ``colour
suppression'' of the electroweak penguin contributions in the decays
$B^0\to\pi^- K^+$ and $B^+\to\pi^+ K^0$ \cite{FM,GR97}, which is
sometimes employed as an argument in favour of their smallness, rests
on a naive cancelation of the contributions of $M_C$ to the sum
$A_{3/2}+A_{1/2}$, which according to (\ref{isofact}) does not take
place in the presence of final-state interactions.

We are now ready to work out the consequences of our results. A
model-independent analysis, which allows for the possibility of having
significant final-state interactions, must assume that the amplitudes
$T$, $C$, $A$ and $P_u$ entering in (\ref{TPA}) all have a similar
magnitude. As discussed above, it must also include the contributions
of electroweak penguin operators. Therefore, we parametrize the isospin
amplitudes in the most general form
\begin{eqnarray}
   \frac{A_{3/2}+A_{1/2}}{\lambda_c(P_c - P_t)}
   &=& - \frac 12 e^{i\gamma} \left( r e^{i\delta} - s e^{i\eta}
    \right) + t e^{i\zeta} \,, \nonumber\\
   \frac{B_{1/2}}{\lambda_c(P_c - P_t)} &=& 1 + \frac 12 e^{i\gamma}
   \left( r e^{i\delta} + s e^{i\eta} \right) \,,
\label{isopar}
\end{eqnarray}
where $r$, $s$ and $t$ are real parameters expected to be of a similar
magnitude, and $\delta$, $\eta$ and $\zeta$ are unknown strong phases.
The electroweak penguin contribution with $\Delta I=0$ is included in
the definition of $P_t$, while that with $\Delta I=1$ defines the term
proportional to $t$. For completeness, we note that in the model
discussed above
\begin{eqnarray}
   r e^{i\delta} &=& e^{i(\phi_{1/2}-\phi_P)}\,
    \left| \frac{M_T}{M_P} \right|\,\left[ 1 + \frac{1+x}{3}
    \left( e^{i\Delta\phi} - 1 \right) \right] \,, \nonumber\\
   - s e^{i\eta} &=& e^{i(\phi_{1/2}-\phi_P)}\,
    \left| \frac{M_T}{M_P} \right|\,\frac{1+x}{3}
    \left( e^{i\Delta\phi} - 1 \right) \,, \nonumber\\
   t e^{i\zeta} &\approx& e^{i(\phi_{1/2}-\phi_P)}\,
    \left| \frac{M_T}{M_P} \right|\,\left[ 0.07 +
    (0.05 + 1.09 x) \left( e^{i\Delta\phi} - 1 \right) \right] \,,
\label{modrel}
\end{eqnarray}
where $x=M_C/M_T\approx 0.2$. It is apparent that rather significant
rescattering effects can arise if the strong phases of the two isospin
amplitudes $A_{1/2}$ and $A_{3/2}$ are different from each other, i.e.\
if $\Delta\phi=O(1)$. The exact theoretical expression for the ratio of
branching ratios in (\ref{FMbound}) becomes
\begin{equation}
   R = \frac{1 + 2r\cos\gamma\cos\delta - 2t\cos\zeta + r^2 + t^2
             - 2rt\cos\gamma\cos(\delta-\zeta)}
            {1 + 2s\cos\gamma\cos\eta + 2t\cos\zeta + s^2 + t^2
             + 2st\cos\gamma\cos(\eta-\zeta)} \,,
\end{equation}
which is the correct generalization of the Fleischer--Mannel result.
Clearly, without additional information about the hadronic parameters
no model-independent bound on the angle $\gamma$ can be derived. A
related question is that about the expected size of the CP asymmetry in
the decays $B^\pm\to\pi^\pm K^0$, for which we find
\begin{eqnarray}
   A_{\rm CP} &=&
   \frac{\mbox{Br}(B^+\to\pi^+ K^0) - \mbox{Br}(B^-\to\pi^-\bar K^0)}
        {\mbox{Br}(B^+\to\pi^+ K^0) + \mbox{Br}(B^-\to\pi^-\bar K^0)}
    \nonumber\\
   &=& - \frac{2s\sin\gamma\,[ \sin\eta + t\sin(\eta-\delta) ]}
          {1 + 2s\cos\gamma\cos\eta + 2t\cos\zeta + s^2 + t^2
           + 2st\cos\gamma\cos(\eta-\zeta)} \,.
\end{eqnarray}
In order to evaluate these results, some information about the
parameters $r$, $s$ and $t$ is required. Model estimates, combined with
the known hierarchy of CKM elements, suggest that $r,s,t=O(0.1)$ (see
also the estimate below, where we show that $|M_T/M_P|=0.14\pm 0.04$),
in which case it is a good approximation to work with the linearized
expressions
\begin{eqnarray}
   R &\approx& 1 + 2\cos\gamma\,(r\cos\delta-s\cos\eta) - 4t\cos\zeta
    \,, \nonumber\\
   A_{\rm CP} &\approx& -2s\sin\gamma\sin\eta \,.
\label{RAapprox}
\end{eqnarray}
{}From these results, it follows that $(R-1)$ and $A_{\rm CP}$ can
naturally be of order 10--20\%, in contrast to claims that the standard
model would not allow for a sizable CP asymmetry in $B^\pm\to\pi^\pm
K^0$ decays \cite{FM,GR97,Kram}. In linear approximation, the CP
asymmetry is insensitive to electroweak penguin contributions. In our
model, using the relations in (\ref{modrel}) and setting $x=0.2$, we
find
\begin{eqnarray}
   R &\approx& 1 + \left| \frac{M_T}{M_P} \right|\,\bigg\{
    2\cos\gamma\,\Big[ 0.2\cos(\phi_{1/2}-\phi_P)
    + 0.8\cos(\phi_{3/2}-\phi_P) \Big] \nonumber\\
   &&\hspace{2.0cm}\mbox{}+ \Big[ 0.8\cos(\phi_{1/2}-\phi_P)
    - 1.1\cos(\phi_{3/2}-\phi_P) \Big] \bigg\} \,, \nonumber\\
   A_{\rm CP} &\approx& 0.8 \left| \frac{M_T}{M_P}
    \right|\,\sin\gamma \left[ \sin(\phi_{3/2}-\phi_P)
    - \sin(\phi_{1/2}-\phi_P) \right] \,.
\label{ACP}
\end{eqnarray}
Note, in particular, the potentially large contribution to $R$ from
electroweak penguins, given by the second term in parenthesis. Unless
$|\phi_{3/2}-\phi_{1/2}|$ is small, this contribution may well dominate
over the term involving the angle $\gamma$. We thus disagree with
Ref.~\cite{FM}, where it was argued that the electroweak penguin
contribution to $R$ is generally very small, i.e.\ of order 1\%, and
can be neglected.

Some information about the magnitude of the isospin amplitudes can be
obtained by employing SU(3) flavour symmetry to relate the $B\to\pi K$
with $B\to\pi\pi$ decay amplitudes. In the process, the CKM parameters
for $b\to s$ transitions have to be replaced by those for $b\to d$
transitions. Since $\lambda_u^{b\to s}/\lambda_u^{b\to d} =
O(\lambda^{-1})$ while $\lambda_t^{b\to s}/\lambda_t^{b\to d} =
O(\lambda)$, where $\lambda=0.22$ is the Wolfenstein parameter, it
follows that electroweak penguin contributions in $B\to\pi\pi$ decays
are much smaller than in $B\to\pi K$ decays. Thus, in the SU(3) limit,
the following triangle relations hold \cite{zepp,Gron}:
\begin{eqnarray}
   3 A_{3/2} &=& {\cal A}(B^+\to\pi^+ K^0)
    + \sqrt{2}\,{\cal A}(B^+\to\pi^0 K^+) \nonumber\\
   &=& {\cal A}(B^0\to\pi^- K^+) + \sqrt{2}\,{\cal A}(B^0\to\pi^0 K^0)
    \nonumber\\
   &=& \frac{V_{us}}{V_{ud}}\,\sqrt{2}\,{\cal A}(B^+\to\pi^+\pi^0)
    + \mbox{electroweak~penguins} \,.
\end{eqnarray}
Using the CLEO measurement $\mbox{Br}(B^+\to\pi^+\pi^0) =
(1.0_{-0.5}^{+0.6})\times 10^{-5}$ \cite{CLEO}, we find that
$|A_{3/2}|_{t=0} = (2.3_{-0.6}^{+0.7})\times 10^{-4}$ (in ``branching
ratio units'', where $|A|=\mbox{Br}^{-1/2}$). This is in good agreement
with our model prediction for the factorized decay amplitudes in
(\ref{MTMC}), which yields $|A_{3/2}|_{t=0} = (2.0\pm 0.5)\times
10^{-4}$ and $|A_{1/2}|_{t=0} = (0.4\pm 0.1)\times 10^{-4}$, where we
have assumed $|V_{ub}| = (3.5\pm 0.5)\times 10^{-3}$. The subscript
``$t=0$'' indicates that these numbers do not include electroweak
penguin contributions. Furthermore, the CLEO measurements
$\mbox{Br}(B^0\to\pi^- K^+) = (1.5_{-0.4}^{+0.5}\pm 0.1\pm 0.1)\times
10^{-5}$ and $\mbox{Br}(B^+\to\pi^+ K^0) = (2.3_{-1.0}^{+1.1}\pm 0.3\pm
0.2)\times 10^{-5}$ \cite{CLEO} imply that $|B_{1/2}-A_{3/2}-A_{1/2}| =
(3.9_{-0.5}^{+0.6})\times 10^{-3}$ and $|B_{1/2}+A_{3/2}+A_{1/2}| =
(4.8_{-1.1}^{+1.3})\times 10^{-3}$. This is a strong indication that
the isospin amplitude $B_{1/2}$, which contains the top- and
charm-penguin contributions, dominates, i.e.\ $|B_{1/2}|\approx (4.1\pm
0.5)\times 10^{-3}\gg|A_{1/2}|$, $|A_{3/2}|$. This information can be
used to estimate the magnitude of one of the hadronic parameters
entering the prediction for the ratio $R$ in (\ref{RAapprox}):
\begin{equation}
   |r\cos\delta - s\cos\eta| < 2\,\left|
   \frac{A_{3/2} + A_{1/2}}{\lambda_c(P_c - P_t)} \right|_{t=0}
   < 2\,\frac{\big(|A_{3/2}| + |A_{1/2}|\big)_{t=0}}{|B_{1/2}|}
   = 0.13\pm 0.03 \,.
\end{equation}
Moreover, in the context of our model, we find from (\ref{isofact})
that
\begin{equation}
   \left| \frac{M_T}{M_P} \right| \approx \frac{3}{1+x}\,
   \frac{|A_{3/2}|_{t=0}}{|B_{1/2}|} = 0.14\pm 0.04 \,.
\end{equation}
These numerical estimates confirm that the effects parametrized by $r$,
$s$ and $t$ are indeed of order 10\%. We thus conclude that the
linearized relations in (\ref{RAapprox}) are reliable.

To summarize, we have argued that the diagrammatic approach to derive
{\em approximate\/} isospin or flavour SU(3) relations between weak
decay amplitudes is misleading in the presence of soft final-state
rescattering effects. There is no theoretical justification to neglect
annihilation topologies, colour-suppressed topologies or up-quark
penguin topologies relative to the colour-allowed tree topology. A
classification scheme based on using isospin amplitudes defined in
terms of operator matrix elements avoids these problems, since all the
above-mentioned amplitudes are contained in the same matrix elements of
the current--current operators $Q_{1,2}^u$, and thus it is natural that
they all have a similar magnitude. On the other hand, many analyses of
CP asymmetries in $B$ decays rely on neglecting such ``suppressed''
contributions. In view of our results, a careful reinvestigation of
these analyses is necessary in order to judge whether amplitude
relations that have been claimed to be ``almost model independent'' are
really theoretically trustworthy. In the present work, we have shown
that the Fleischer--Mannel bound for $\gamma$ \cite{FM} relies on such
unjustified assumptions. With a realistic treatment of final-state
interactions, no useful bound can be obtained. In particular, we have
shown that the effects of electroweak penguins are severely enhanced in
the presence of different strong phases for the isospin amplitudes
$A_{1/2}$ and $A_{3/2}$ and, in principle, can yield a contribution to
$R$ of as much as 10--20\%. We have also shown that, in the context of
the standard model, the CP asymmetry in $B^\pm\to\pi^\pm K^0$ decays
can be of order 10\%.

The problem that rescattering effects could invalidate the results
derived using a diagrammatic amplitude analysis has been pointed out
previously by Wolfenstein \cite{Wolf} and by Soni \cite{Soni}. The
issue has also been discussed recently in Ref.~\cite{BFM}, where
conclusions different from ours have been reached. In that paper, the
authors perform an isospin analysis and absorb possible soft
rescattering contributions into the amplitudes $T$ and $P$,
corresponding to the choice $\Delta P=-X$ in our notation in
(\ref{Xcontr}). However, then they neglect these effects by implicitly
assuming that the rescattering contribution $\Delta P$ to the up-quark
penguin is much smaller than the tree amplitude $T$. As we have shown,
this assumption is not justified.

While this paper was in writing, we became aware of a letter by
G\'erard and Weyers \cite{GeWe}, in which conclusions similar to ours
are reached. In particular, in a model with quasi-elastic rescattering
these authors derive the first two relations in (\ref{model}).

\vspace{0.3cm}
{\it Acknowledgments:\/}
My interest in this subject was initiated in lively discussions with
A.~Kagan, Y.~Nir, R.~Fleischer and T.~Mannel. I would like to thank
B.~Stech for useful comments.


\begin{thebibliography}{99}
\parskip=0pt

\bibitem {Flei}
R. Fleischer, Int.\ J.\ Mod.\ Phys.\ A {\bf 12}, 2459 (1997).

\bibitem {rev}
Y. Grossman, Y. Nir and R. Rattazzi, Preprint SLAC-PUB-7379
[hep-ph/9701231], to appear in the Second Edition of {\sl Heavy
Flavours}, edited by A.J. Buras and M. Lindner (World Scientific,
Singapore);
Y. Nir, Preprint WIS-97/28/Sep-PH [hep-ph/9709301], to appear in the
Proceedings of the 18th International Symposium on Lepton Photon
Interactions, Hamburg, Germany, July 1997.

\bibitem {zepp}
D. Zeppenfeld, Z. Phys.\ C {\bf 8}, 77 (1981).

\bibitem {GL}
M. Gronau and D. London, Phys.\ Rev.\ Lett.\ {\bf 65}, 3381 (1990).

\bibitem {MGr}
M. Gronau, Phys.\ Lett.\ B {\bf 265}, 389 (1991).

\bibitem {NQ}
Y. Nir and H.R. Quinn, Phys.\ Rev.\ Lett.\ {\bf 67}, 541 (1991);\\
H.J. Lipkin, Y. Nir, H.R. Quinn and A.E. Snyder, Phys.\ Rev.\ D {\bf
44}, 1454 (1991).

\bibitem {Chau}
L.L. Chau et al., Phys.\ Rev.\ D {\bf 43}, 2176 (1991).

\bibitem {Gron}
M. Gronau, J.L. Rosner and D. London, Phys.\ Rev.\ Lett.\ {\bf 73}, 21
(1994);
O.F. Hern\'andez, D. London, M. Gronau and J.L. Rosner, Phys.\ Lett.\ B
{\bf 333}, 500 (1994); Phys.\ Rev.\ D {\bf 50}, 4529 (1994).

\bibitem {Heff}
For a review, see:
G. Buchalla, A.J. Buras and M.E. Lautenbacher, Rev.\ Mod.\ Phys.\ {\bf
68}, 1125 (1996).

\bibitem {BuFl}
A.J. Buras and R. Fleischer, Phys.\ Lett.\ B {\bf 341}, 379 (1995).

\bibitem {charmp}
M. Ciuchini, E. Franco, G. Martinelli and L. Silvestrini, Nucl.\ Phys.\
B {\bf 501}, 271 (1997);
M. Ciuchini et al., Preprint CERN-TH/97-188 [hep-ph/9708222].

\bibitem {FM}
R. Fleischer, Phys.\ Lett.\ B {\bf 365}, 399 (1996);\\
R. Fleischer and T. Mannel, Preprint TTP97-17 [hep-ph/9704423].

\bibitem {CLEO}
CLEO Collaboration (R. Godang et al.), Preprint CLNS~97-1522 (1997);\\
F. W\"urthwein (representing the CLEO Collaboration), [hep-ex/9706010],
to appear in the Proceedings of Les Rencontres du Moriond: {\sl QCD and
High Energy Hadronic Interactions}, Les Arcs, France, March 1997.

\bibitem {GNPS}
Y. Grossman, Y. Nir, S. Plaszczynski and M.H. Schune, Preprint
SLAC-PUB-7622 [hep-ph/9709288].

\bibitem {GR97}
M. Gronau and J.L. Rosner, Preprint CALT-68-2142 [hep-ph/9711246].

\bibitem {Dono}
J.F. Donoghue, E. Golowich, A.A. Petrov and J.M. Soares, Phys.\ Rev.\
Lett.\ {\bf 77}, 2178 (1996).

\bibitem {Stech}
M. Neubert and B. Stech, Preprint CERN-TH/97-99 [hep-ph/9705292], to
appear in the Second Edition of {\sl Heavy Flavours}, edited by A.J.
Buras and M. Lindner (World Scientific, Singapore).

\bibitem {Kram}
G. Kramer, W.F. Palmer and H. Simma, Z.~Phys.\ C {\bf 66}, 429 (1995).

\bibitem {Wolf}
L. Wolfenstein, Phys.\ Rev.\ D {\bf 52}, 537 (1995).

\bibitem {Soni}
A. Soni, in {\sl Penguins, charmless $B$ decays and CP violation}, to
appear in the Proceedings of the International Europhysics Conference
on High Energy Physics, Jerusalem, Israel, August 1997.

\bibitem {BFM}
A.J. Buras, R. Fleischer and T. Mannel, Preprint CERN-TH/97-307
[hep-ph/9711262].

\bibitem {GeWe}
J.M. G\'erard and J. Weyers, Preprint UCL-IPT-97-18 [hep-ph/9711469].

\end{thebibliography}
\end{document}